# DeepRegularizer: Rapid Resolution Enhancement of Tomographic Imaging using Deep Learning

DongHun Ryu, Dongmin Ryu, YoonSeok Baek, Hyungjoo Cho, Geon Kim, Young Seo Kim, Yongki Lee, Yoosik Kim, Jong Chul Ye, Hyun-Seok Min, and YongKeun Park

*Abstract*—Optical diffraction tomography measures the three-dimensional refractive index map of a specimen and visualizes biochemical phenomena at the nanoscale in a non-destructive manner. One major drawback of optical diffraction tomography is poor axial resolution due to limited access to the three-dimensional optical transfer function. This missing cone problem has been addressed through regularization algorithms that use a priori information, such as non-negativity and sample smoothness. However, the iterative nature of these algorithms and their parameter dependency make real-time visualization impossible. In this article, we propose and experimentally demonstrate a deep neural network, which we term DeepRegularizer, that rapidly improves the resolution of a three-dimensional refractive index map. Trained with pairs of datasets (a raw refractive index tomogram and a resolution-enhanced refractive index tomogram via the iterative total variation algorithm), the three-dimensional U-net-based convolutional neural network learns a transformation between the two tomogram domains. The feasibility and generalizability of our network are demonstrated using bacterial cells and a human leukaemic cell line, and by validating the model across different samples. DeepRegularizer offers more than an order of magnitude faster regularization performance compared to the conventional iterative method. We envision that the proposed data-driven approach can bypass the high time complexity of various image reconstructions in other imaging modalities.

## I. INTRODUCTION

OPTICAL diffraction tomography (ODT) has emerged as a powerful label-free imaging technique in biomedicine[1-3]. The technique reconstructs the three-dimensional (3D) refractive index (RI) map of a biological specimen at the nanoscale without any labelling agent. Although the widespread versatility of ODT has been demonstrated in numerous research areas [4], e.g. cell biology [5-7], microbiology [8], haematology [4, 9], and nanoscience [7, 10], one major drawback must be circumvented to further expand the applicability of ODT: its relatively poor axial resolution [11]. Because of the limited access to scattered fields, the scattering potential of a sample along the optical axis is highly underestimated [12]. This leads to poor axial resolution and inaccurate reconstruction of RI maps. The inaccessible region surrounding the optical axis in the 3D optical transfer function generates streaking and elongated tail artefacts in the reconstructed tomogram (Fig. 1 and Fig. S1). A wide variety of efforts, from optical to numerical, have been made to address this 'missing cone problem.' The missing cone has been treated by rotating the sample [4, 13-15] or engineering illumination [16], but these methods require highly stable optical systems and strict sample conditions.

Computational strategies imposing specific constraints, such as non-negativity, edge-preserving smoothness, and total variation, have also been used to fill in the missing cone in tomographic imaging [17-22]. However, these methods suffer from high computational cost because their algorithms iteratively search for a solution; the processing time scales from minutes to hours to even days, depending on the sample size, constraints, and computational power [17, 23-25]. Optimization-based diffraction tomography that combines the inverse problem with various regularizers in the reconstruction pipeline tends to be more computationally demanding with the use of proximal gradients [24-27].

Deep learning has rapidly become an attractive tool for optical imaging, spanning a wide range of image sensing and

Manuscript received xxx xx, 2020.This work was supported in part by KAIST, Tomocube, and National Research Foundation of Korea (2015R1A32066550, 2017M3C1A3013923, 2018K000396) (Corresponding authors: YongKeun Park; Hyun-Seok Min).

D. H. Ryu, Y. S. Baek, G. Kim, and Y. S. Kim are with the Department of Physics, and KAIST Institute for Health Science and Technology, Korea Advanced Institute of Science and Technology, Daejeon 34141, Republic of Korea (dryu@kaist.ac.kr; lovebaek@kaist.ac.kr; gyugkei@kaist.ac.kr; ykim717@kaist.ac.kr).

D. Ryu, H. Cho, and H.-S. Min are with Tomocube Inc., Daejeon 34051, Republic of Korea (dmryu@tomocube.com; hjcho@tomocube.com; hsmin@tomocube.com).

Y. Lee and Y. Kim are with the Department of Chemical and Biomolecular Engineering, and KAIST Institute for Health Science and Technology, Korea Advanced Institute of Science and Technology, Daejeon 34141, Republic of Korea (luckily@kaist.ac.kr; ysyoosik@kaist.ac.kr).

J. C. Ye is with the Department of Bio and Brain Engineering, and the Department of Mathematical Sciences Korea Advanced Institute of Science and Technology, Daejeon 34141, Republic of Korea (jong.ye@kaist.ac.kr).

Y. K. Park is with with the Department of Physics, and KAIST Institute for Health Science and Technology, Korea Advanced Institute of Science and Technology, Daejeon 34141, Republic of Korea, and also with Tomocube Inc., Daejeon 34051, Republic of Korea (yk.park@kaist.ac.kr).

D. H. Ryu and D. Ryu contributed equally to this work.

restoration problems [28-32], such as super-resolution imaging [33, 34], imaging through turbid media [35-38], quantitative phase imaging [39-41], denoising [42, 43], digital staining [44-46], and single-pixel imaging [47]. These deep learning-based image restorations are realized by training a network that learns the underlying forward/inverse operation of a given optical system or specific image-to-image transformation. Once the network is optimally trained, it can rapidly perform an image inference task without any iterations or additional parameter tuning.

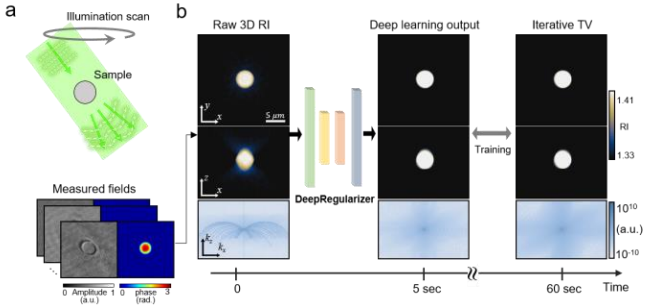

Fig. 1. ODT and regularization process using our 3D deep neural network (DeepRegularizer) and conventional iterative TV algorithm. (a) Schematic of ODT where scattered fields of a microbead at various angles are used to reconstruct the 3D RI via the Fourier diffraction theorem (See Methods and Materials section). (b) The reconstructed RI tomogram, suffering from severe artefacts caused by the missing cone problem along kz, is regularized via DeepRegularizer in 5 sec, compared to approximately 1 min by the conventional iterative TV optimization exploiting the split Bregman iteration

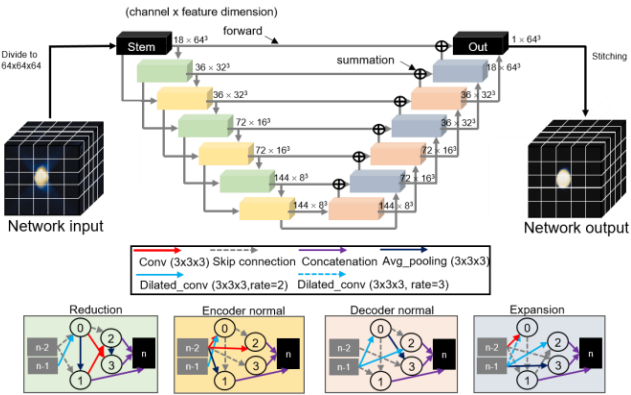

Fig. 2. Architecture of DeepRegularizer. Our trained 3D deep neural network regularizes a patch of tomogram (64 × 64× 64) and stitches all regularized patches into the entire tomogram in a few seconds. Four cells (reduction, encoder normal, decoder normal, and expansion) are alternatively used in the downsampling and upsampling paths. Each cell consists of a 3D convolution operation (64 × 64× 64), skip connection, concatenation, average pooling (3 × 3× 3), and dilated convolution (3 × 3× 3, rate = 2 or 3); the standard convolution is used for the reduction and encoder normal cell in the downsampling path and the dilated convolution is used for the decoder normal and expansion cell in the upsampling path to preserve feature information and efficiently expand feature dimension.

Here we propose a deep learning-based regularization framework that rapidly improves the resolution of a 3D RI tomogram, termed DeepRegularizer (Fig. 1). To overcome the high computational cost of the total variation (TV) regularization widely used in tomographic imaging, we train a 3D U-net-based convolutional neural network (Fig. 2). This neural network learns a translation between a raw tomogram and its regularized tomogram as the pairs of a dataset obtained from the conventional TV algorithm. We experimentally verify the performance of DeepRegularizer with bacteria (*Streptococcus pyogenes*) and a human acute myeloid leukemia cell line (OCI-AML3). We also demonstrate that this approach has a strong generalizability by cross-validating across two sample types and by inferring on various suspension and adherent cell types. The architecture of the proposed neural network is designed and optimized by exploiting a recent auto machine learning technique, termed scalable neural architecture search (ScNAS) [48], which significantly alleviates laborious architecture optimization. The proposed computational framework, which requires minimal optimization in minor hyperparameters such as batch size, can be utilized in various optical imaging tasks and relaxes the high computational costs of image reconstruction problems.

## II. METHODS AND MATERIALS

### A. DeepRegularizer Implementations

Our 3D U-net-based network that learns a mapping function between a raw tomogram and a TV regularized tomogram was constructed based on a ScNAS [49], which significantly alleviated a routine hyper-parameter search and an exhaustive architecture search. The ScNAS utilizes a stochastic sampling algorithm for the bi-level optimization of network parameters. The details of the ScNAS are beyond the scope of this paper and are rigorously explained in refs [49, 50].

The architecture of our deep neural network is shown in Fig. 2. As demonstrated in various image processing tasks, our U-net-based network also consists of downsampling and upsampling paths. In the downsampling path, which alternates a reduction cell and an encoder normal cell, smaller features with an increasing number of channels are extracted and also propagated to the corresponding level of the upsampling path for combining and refining spatial information in high resolution. In the downsampling path, each cropped 64 × 64 × 64 patch of tomogram is processed to $18 \times 64^3$, $36 \times 32^3$, $72 \times 16^3$, and $144 \times 8^3$, passing through a stem cell and alternatively the reduction and encoder normal cell. The stem cell simply expands the channel of the cropped tomogram to $18 \times 64^3$ by performing convolution, ReLu, global average pooling, linear, ReLu, linear, sigmoid, and instance norm in series. Each cell takes the output of two consecutive previous cells as an input consisting of a 3D convolution (3 × 3 × 3), a skip connection, concatenation, average pooling, and a dilated 3D convolution (3 × 3 × 3, rate = 2 or 3) with concurrent directions. This helps simultaneously extract the fine-detail spatial information and global text of a 3D tomogram. In the upsampling path, the high-resolution features from the downsampling and extracted information in this expansion are combined via an alternating sequence of decoder normal and expansion cell. Note that the dimension of the extracted features increases while the number of channels shrinks. It is also noteworthy that the cells in the upsampling path are mainly comprised of the dilated convolution in order to preserve the feature resolution and efficiently expand its dimension. Finally, a regularized tomogram with a single channel (1 × 64³), which of the dimension is identical to the input tomogram dimension, is generated by a single convolution and stitched into the entire tomogram using the processing algorithm explained in a later section.

The network was implemented in Pytorch 1.0 using a GPU server computer (Intel® Xeon® CPU E5-2620 v4 and 8 of NVIDIA GeForce GTX 1080 Ti). We trained our network using Adam optimizer (learning rate = 0.0001) with a $l_2$ loss. Though the learning capacity of our 3D U-net trained with the $l_2$ loss seems sufficient enough to output such significantly small error levels compared to the ground truth pair, various metrics, such as the structural similarity index could be employed for further optimization. The learnable parameters were initialized by He initialization. We augmented data using flip and rotation with a rate of 0.3. We trained our algorithm with a batch of 80 for approximately 24 hours, yet the time highly depended on computing resources. We selected our best model at 200 epochs and monitored validation accuracy based on MSE.

### B. Tomogram Stitching

Because we cropped each raw tomogram into $64^3$ voxel-patches for regularization, we had to stitch the regularized chunks into the original size of the tomogram. As each cropped tomogram overlapped one another for 32 × 64 × 64 voxels with a stride of 32, we summed up neighbouring patches and accommodated for the overlapped voxels using a spline window widely used to process an overlap region smoothly. Then, we divided by 8 to normalize the overlapped voxels. The resultant regularized tomogram using DeepRegularizer has the identical size as the raw tomogram.

### C. Optical Diffraction Tomography

The RI tomograms of OCI-AML3 cell were acquired using a commercialized ODT system outfitted with a beam-scanning digital micromirror device (DMD) based on Mach-Zehnder interferometry (HT-2H, Tomocube Inc., Republic of Korea). The optical setup is shown in Fig. S2. A diode-pumped solid-state laser beam (532-nm wavelength, 10 mW, MSL-S-532-10 mW, CNI laser, China) is split into a sample beam and reference beam through a 1×2 fibre coupler (OZoptics, Canada). The sample beam, spatially modulated by the DMD (DLP65300FYE, Texas Instruments, USA), impinges upon a sample after passing through a condenser objective lens (×60, Numerical aperture (NA) = 1.2). Then, the scattered light is transmitted through an imaging objective lens (×60, NA = 1.2) and combined with the reference beam to form an interferogram at the CMOS camera (FL3-U3-13Y3M-C, FLIR Systems, USA) after passing a linear polarizer. In total, forty-nine different angles of interferograms are measured.

Subsequently, a 3D tomogram is reconstructed from the measured interferograms, which were all processed in MATLAB (MathWorks, USA). First, 2D optical fields at the sample plane are retrieved from each captured interferogram using a field retrieval algorithm exploiting spatial filtering [51]. Then, following the Fourier diffraction theorem that governs a relation between 2D optical fields and sample RI with Rytov assumption [52], the 3D RI tomogram of the sample is reconstructed. The lateral and axial resolution, owing to Lauer criteria [2], are 110 nm and 355 nm, respectively. The time to acquire the interferograms is 0.1 seconds and it typically takes a few seconds to reconstruct the tomogram from the interferograms using a standard personal computer, depending on the voxel size.

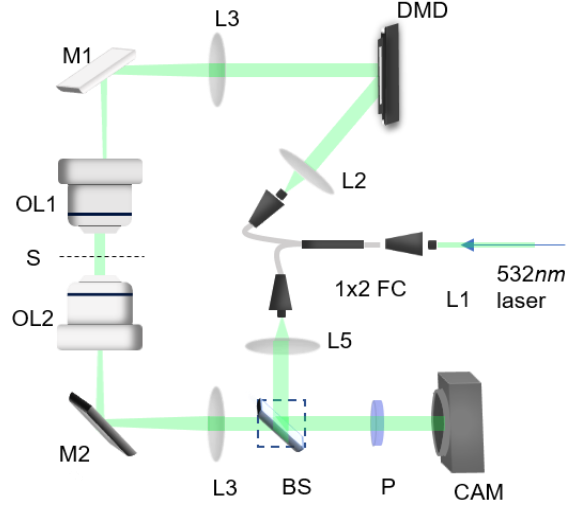

Fig. 3. Optical setup of the imaging system. L: Lens, FC: Fibre coupler, DMD: Digital micromirror device, BS: Beam splitter, M: Mirror, OL: Objective lens, S: Sample, P: Linear polariser, and CAM: Camera.

For the *S. pyogenes* dataset, we measured tomograms from a different ODT imaging system. The principal components were identical to the explained system (Fig. 3), except for the number of illumination angles (equal to 71) and the two objective lenses (×60, NA = 0.8), which results in relatively lower lateral and axial resolution (166 nm and 1000 nm).

### D. Total Variation Regularization with a Non-negativity Constraint

To generate the paired dataset, obtained raw RI tomograms are processed using an iterative regularization algorithm with a TV and non-negativity constraint. We employed the split Bregman method to implement fast regularization, which has been widely used for image processing tasks. First, one can pose the regularization problem as an unconstrained minimization problem:

$$\min_f E(f) \text{ such that } Af = g \quad (1)$$

where $A$, $f$, $g$, and $E(f)$ are the Fourier transform operator, scattering potential, measured Fourier transformed scattering potential, and an energy functional consisting of TV, respectively, and $T(f) = \sqrt{(\nabla_x f)^2 + (\nabla_y f)^2 + (\nabla_z f)^2}$ and the non-negativity functionals are $N(f)$.

To solve the constrained problem in a simple manner, the optimization problem can be formulated in an unconstrained problem as follows:

$$\|Af - g\|_2^2 + T(f) + N(f \geq 0). \quad (2)$$

Then, the split Bregman framework is employed to solve this optimization problem as suggested by Goldstein, which can be re-formulated as follows:

$$f^{k+1} = \min_f \| Af - g \|^2_2 + \sqrt{(\nabla_x f)^2 + (\nabla_y f)^2 + (\nabla_z f)^2} + N(f \geq 0) \quad (3)$$

$$g^{k+1} = g^k + g - Af^{k+1}. \quad (1)$$

Finally, variable splitting is used to solve (3) and we implemented the iterative regularization algorithm that follows in MATLAB (MathWorks), using a GPU server computer (Intel® Xeon® CPU E5-2620 v4 and 8 of NVIDIA GeForce GTX 1080 Ti). For fair inference test, we used a single GPU to compare our deep learning algorithm and the iterative regularization algorithm, although all 8 GPUs were used for deep learning training. The derivation and algorithm of the split Bregman framework are described in Appendix B.

Input: $N, M, \mu, \tau, \gamma$
Initialize:
$f^0 = A^T g$, $b_x^0 = b_y^0 = b_z^0 = b_w^0 = d_x^0 = d_y^0 = d_z^0 = w^0 = 0$.
For OuterIter = 1 to N
  For InnerIter = 1 to M

$$f^{k+1} = [\mu A^T A + \lambda \Delta + \gamma I]^{-1} rhs^k$$

$$d_x^{k+1} = \max(s^k - \frac{1}{\lambda}, 0) \frac{\nabla_x f^{k+1} + b_x^k}{s^k}$$

$$d_y^{k+1} = \max(s^k - \frac{1}{\lambda}, 0) \frac{\nabla_y f^{k+1} + b_y^k}{s^k}$$

$$d_z^{k+1} = \max(s^k - \frac{1}{\lambda}, 0) \frac{\nabla_z f^{k+1} + b_z^k}{s^k}$$

$$w^{k+1} = \max(| f^{k+1} + b_w^k | - \frac{1}{\gamma}, 0) \frac{f^{k+1} + b_w^k}{| f^{k+1} + b_w^k |}$$

$$b_x^{k+1} = b_x^k + (\nabla_x^T f^{k+1} - d_x^{k+1})$$
$$b_y^{k+1} = b_y^k + (\nabla_y^T f^{k+1} - d_y^{k+1})$$
$$b_z^{k+1} = b_z^k + (\nabla_z^T f^{k+1} - d_z^{k+1})$$
$$b_w^{k+1} = b_x^k + (\nabla_x^T f^{k+1} - w^{k+1})$$
$$g^{k+1} = g^k + g - Af^{k+1}$$

Where
$$rhs^k = [\mu A^T g + \lambda(\nabla_x^T (d_x - b_x) + \nabla_y^T (d_y - b_y) + \nabla_z^T (d_z - b_z)) + \gamma(w - b_w)]$$
$$\Delta = -(\nabla_x^T \nabla_x + \nabla_y^T \nabla_y + \nabla_z^T \nabla_z)$$

The five algorithmic parameters, $N, M, \mu, \tau,$ and $\gamma$, were extensively grid-searched and manually selected for the generation of the paired dataset. Because we do not have a ground truth for experimental data, the magnitude of the parameters must be manually tuned to optimize reconstruction quality while attempting to sharpen sample boundary sharpness and conserve subcellular features. The optimized parameters ($N, M, \mu, \tau,$ and $\gamma$) for the beads, *S. pyogenes*, and OCI-AML3 cell are (2, 400, 10, 10, and 1), (5, 100, 50, 50, and 1), and (3, 60, 150, 150, and 1), respectively.

### E. Sample Preparation

Bacteria samples for tomographic imaging were obtained from a laboratory culture of *S. pyogenes*. The frozen glycerol stocks of bacteria, stored at -80 °C, were slowly defrosted at room temperature. The bacteria were separated from the glycerol solution by repeated centrifugation and washing with fresh medium. After stabilizing in a 35 °C incubator, the bacteria were streaked on agar plates. The agar plates were incubated at 35 °C until colonies were visible to the naked eye. Single colonies were inoculated on fresh medium using a sterile loop. Each subculture was incubated in a 35 °C shaking incubator until the concentration reached approximately 108-109 cfu/ml. The grown subculture was repeatedly centrifuged and washed, once with fresh medium and twice with phosphate-buffered saline solution. Finally, the washed bacteria solution was diluted with phosphate-buffered saline solution until the concentration was adequate for single-sample imaging.

The human acute leukemic cells (OCI-AML3) were obtained from Seoul National University Hospital. After the cells in the cryovial were promptly thawed at room temperature in a 37 °C water bath and centrifuged at 15000 RPM, the cells were resuspended and cultured in a fresh medium (alpha-minimum essential medium (Welgene, Korea) with 20% fetalgro bovine growth serum (RMbio, USA)). Note that the processed cells were incubated at 37 °C in a humidified atmosphere of 5 % $CO_2$ and sub-cultured every three days with a proper density (~$10^6$ to $2 \times 10^6$ cells/ml) to maintain optimal conditions before imaging. Finally, the sample was diluted with the culture medium until the concentration was approximately $2.5 \times 10^5$ cells / mL for single-sample imaging.

Two hundred and seventeen samples of *S. pyogenes* and 614 samples of OCI-AML3 were imaged using the tomographic microscope. The acquired tomograms were regularized using the regularization solver described in the previous section. Since we do not have a ground truth, the parameters were manually tuned and the visualized tomograms were checked at various perspectives, e.g. *x-y* and *x-y* slice images and maximum intensity projection images.

### III. RESULTS

We experimentally verified our method for tomogram regularization of biological samples and compared it to iterative TV regularization using split Bregman-based optimization. This optimization is regarded as the ground truth for deep learning training. For the results in the subsequent sections, our deep neural network was trained with dataset pairs and tested on unseen tomogram data not used in training. First, we imaged and regularized *S. pyogenes* and the OCI-AML3 cells for the experimental demonstration of DeepRegularizer (III-A). Next, we present cross-sample validation results to demonstrate the generalizability of DeepRegularizer. The neural network trained with bacteria tomograms was tested on myeloid leukemic cell tomograms with entirely different data distributions; reversely the network that learned the myeloid leukemic cell data distribution was also tested on the other dataset (III-B). For additional generalization test, we regularized test dataset, including MV4-11, monocyte, H292, and NIH3T3 using the model trained with the leukemic cell (Appendix C). Finally, we compared the computation time of our rapid regularizer with that of the iterative TV algorithm (III-C).

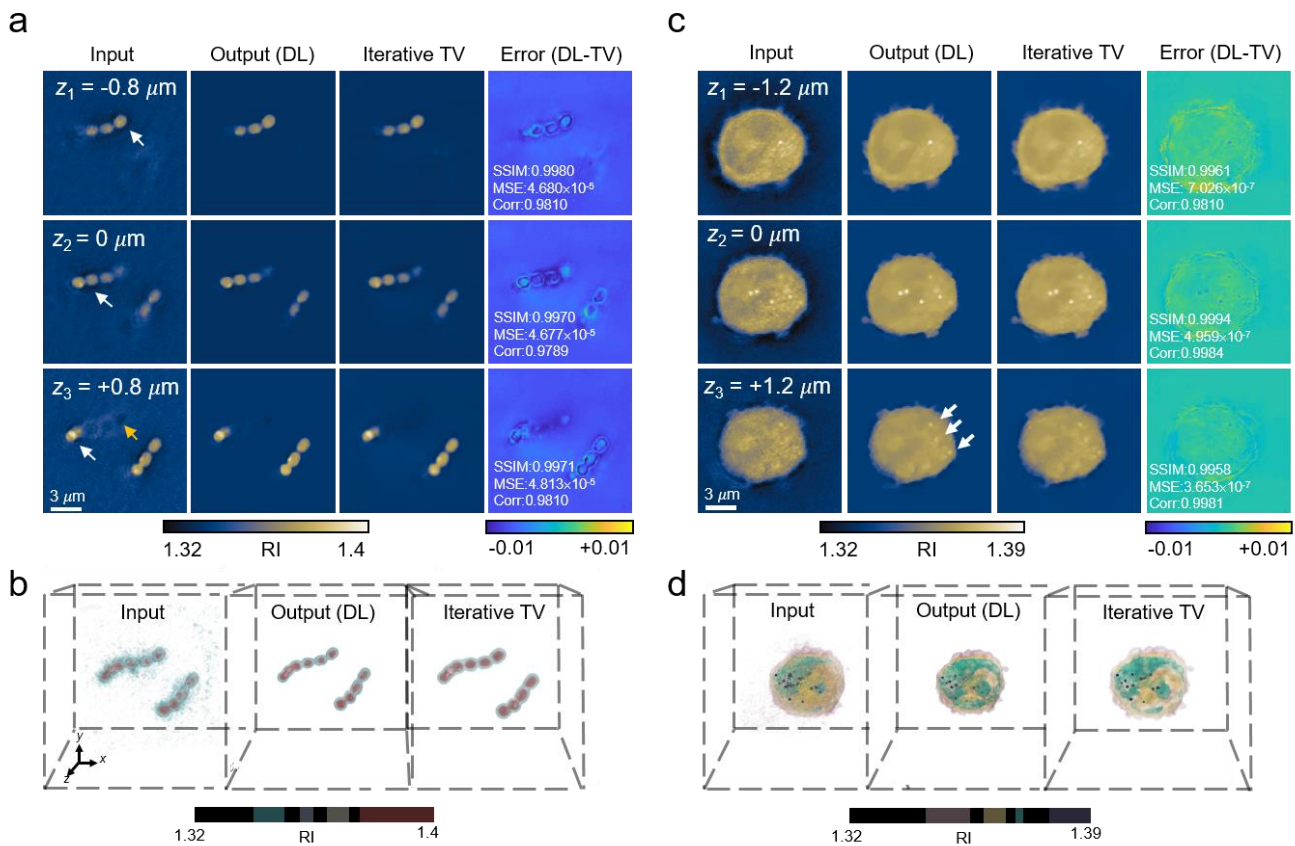

Fig. 4. Experimental demonstration of DeepRegularizer. Our deep learning results compared to the iterative TV algorithm for the regularization of S. pyogenes tomograms (a-b) and OCI-AML3 tomograms (c-d) are shown. (a,c) Three lateral slices for the input, DL output, and iterative TV are shown with error maps and their quantifications using three metrics (structure similarity index, mean-square-error, and Pearson correlation coefficient). White arrows in a indicate the blurring-out artefact in the raw tomogram of S. pyogenes. Remaining features interfering with other lateral planes in the tomogram are indicated by yellow arrows in a. White arrows in c show that our DeepRegularizer reveals granules that are not clearly visible in the input tomogram. (b,d) The corresponding iso-surface rendered images are visualized.

### A. 3D Resolution Enhancement of Biological Samples

We first demonstrate the performance of DeepRegularizer with bacteria (*S. pyogenes*), visualizing three different *x-y* slices ($z_1 = -0.8\ \mu m$, $z_2 = 0\ \mu m$, and $z_3 = +0.8\ \mu m$) and an iso-surface 3D rendering of corresponding tomograms (Fig. 4a-b). In both results, DeepRegularizer shows improved resolution and the non-sporing cocci of two *S. pyogenes* are better resolved compared to the input tomograms. The raw tomogram exhibits halo effects near the boundary of *S. pyogenes* and ring-shaped artefacts, which are highlighted with white and orange arrows, respectively. DeepRegularizer not only corrects the artefacts but also suppresses the RI fluctuations in the background, making the edges of bacteria cocci sharp (Fig. 4a, output). The network output also shows a good agreement with the iterative TV regularization (Fig. 4a, iterative TV). To quantify the results of DeepRegularizer against the iterative TV algorithm, we evaluate the structural similarity index (SSIM), mean square error (MSE), and Pearson correlation coefficient (Corr) along with each error map. SSIM for $z_1$, $z_2$, and $z_3$ are 0.9980, 0.9970, and 0.9971, respectively; MSE for $z_1$, $z_2$, and $z_3$ are $4.680 \times 10^{-5}$, $4.677 \times 10^{-5}$, and $4.813 \times 10^{-5}$, respectively; Pearson correlation coefficients for $z_1$, $z_2$, and $z_3$ are 0.9810, 0.9789, and 0.9810, respectively, validating the high-fidelity of the DeepRegularizer framework. The detailed subcellular structure of *S. Pyogenes* is more perspicuously visualized in the 3D rendering (Fig. 4b).

To further demonstrate DeepRegularizer, we used the OCI-AML3 cells, which contain more complex subcellular compartments. The 3D morphological appearance of nucleus and granules are improved via DeepRegularizer and the unwanted noises, spread out in the 3D RI map, are removed from the input tomogram. This is illustrated in three *x-y* slice images and benchmarked against the iterative TV results (Fig. 4c). Through the DeepRegularizer framework, the halo artefacts surrounding the cell boundary are significantly reduced, clearly visualizing membrane curvatures of OCI-AML3 cells. Our framework also reveals the granules inside the cytoplasm that are not resolvable by the input tomogram, as indicated by arrows (Fig. 4c, $z_3$ = 1.2 $\mu m$). While some of the regions, e.g. edges of the cell boundary or subcellular features, remain comparatively uncorrelated as shown in the error maps, the robust 3D resolution enhancement performance of DeepRegularizer is verified by the metrics given earlier (SSIM, MSE, and Corr). The 3D rendering of the OCI-AML3 tomogram also validates the resolution improvement in our framework, providing the definite nucleus and granules inside the cells (Fig. 4e).

### B. Generalizability: Cross-sample Validation

To test the generalizability of DeepRegularizer, we conducted a performance test where the training data distribution differed from the test data distribution. First, the model trained by the OCI-AML3 dataset regularizes the *S. pyogenes* dataset (Fig. 5a, column 1). The result is comparable to the inference of the model trained with the *S. pyogenes* (Fig. 5a, column 2) and visualized in maximum intensity projection images (MIP). The cross-validation test where the *S. pyogenes* model is applied to OCI-AML3 also shows a successful reconstruction (Fig. 5b, column 1) consistent with the result of the OCI-AML3 model (Fig. 5b, column 2). The error maps, along with the three metrics (SSIM for bacteria and OCI-AML3: 0.9982 and 0.9979; MSE for bacteria and OCI-AML3: 1.3232 × $10^{-5}$ and 1.3673 × $10^{-5}$; Pearson correlation coefficient for bacteria and OCI-AML3: 0.9961 and 0.9980) comparing the two different training models, further demonstrate the strong generalization capability and visualize the negligible error levels of the regularized tomograms inferred by the two different models with separate data distributions.

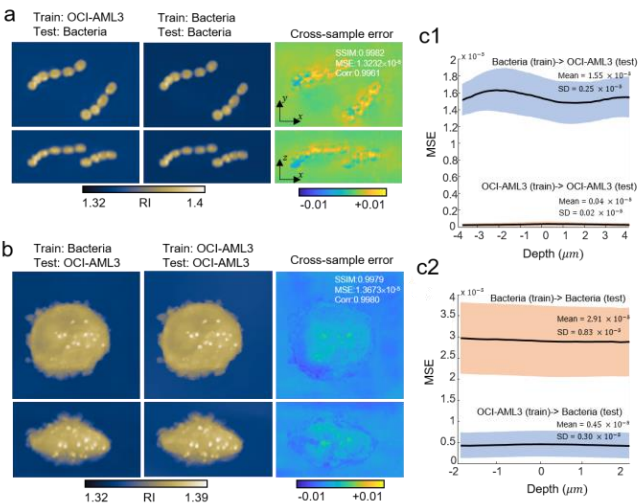

Fig. 5. Generalizability of DeepRegularizer. Inference in significantly different data distributions that are not used during training is illustrated: a Test on bacteria dataset using the model trained with the OCI-AML3 dataset and b Test on the OCI-AML3 dataset using the model trained with the bacteria dataset. The corresponding results where the training and test dataset have the same distribution are compared with the error maps and their three evaluation metrics. c1-c2 Quantification of two different scenarios (a-b) using MSE. The MSE for each lateral slice of the inferred tomogram is computed against the ground truth training dataset (iterative TV). Considering the axial dimension of a tomogram, different depths are investigated for the WBC (-4 $\mu m$ to +4 $\mu m$) and bacteria (-2 $\mu m$ to +2 $\mu m$).

To quantitatively assess the cross-sample performance for the entire test dataset, we computed and compared MSE of each *x-y* plane for two scenarios: (1) where training and test set have the same data distribution and (2) where a model trained using a data type is tested on a different type of dataset. For the entire test set of *S. pyogenes* (*n* = 162) and OCI-AML3 (*n* = 100), test tomograms were examined for ±2 $\mu m$ (*S. pyogenes*) or ±4 $\mu m$ (OCI-AML3) from the focal plane. For the test set of OCI-AML3, scenario (1) (mean: 0.04 × $10^{-5}$, standard deviation (SD): 0.02 × $10^{-5}$) outperformed scenario (2) (mean: 1.55 × $10^{-5}$, SD: 0.25 × $10^{-5}$), as illustrated in Fig. 5c1. This is a common case for a data-driven approach as the network that approximates a non-linear mapping between two image domains tends to cope better when the test set and training set come from the same data distribution. On the contrary, when the test set is *S. pyogenes*, the model trained by the OCI-AML3 (mean value: 0.45 × $10^{-5}$, SD: 0.30 × $10^{-5}$) overtook the model trained by the same dataset (mean value: 2.91 × $10^{-5}$, SD: 0.83 × $10^{-5}$, Fig. 5c2). This result implies that the DeepRegularizer framework can learn diverse representations of the 3D image transformation, enabling powerful resolution enhancement tasks that do not overfit specific data distributions. We believe this is due to the fact that the deep learning model trained with OCI-AML3 has higher information capacity that encompasses the features of *S. pyogenes*, leading to superior performance compared to the model trained with *S. pyogenes*. Moreover, the generalizability across different imaging devices is verified to a certain degree because the two datasets were obtained from two independent ODT systems and different experimental protocols that are further detailed in the Materials and Methods section.

### C. Inference Time: DeepRegularizer versus iterative TV

To validate our rapid resolution enhancement capability, we compared the running time of DeepRegularizer and a conventional TV regularizer based on the voxel size of an input tomogram (Fig. 6). We changed the dimension of *x* and *y*, and fixed the *z*-dimension as 64, $64^3$, $96^2 \times 64$, $128^2 \times 64$, $160^2 \times 64$, $192^2 \times 64$, $224^2 \times 64$, $256^2 \times 64$, $288^2 \times 64$, and $320^2 \times 64$. We measured the time to obtain a 3D regularized tomogram from a raw tomogram for both algorithms. For DeepRegularizer, which can process a patch of $64^3$ voxels, running time also includes data processing times, such as patching and stitching, whereas the iterative TV can input and output the whole 3D tomogram. For a fair comparison, we fixed two iterations for the conventional TV (inner iterations: 100; outer iterations: 5) as suggested by the work of Osher et al [23] that stated the aforementioned numbers would be empirically sufficient for the convergence of the split Bregman method for L1-regularized problems.

The results show that our approach is more than an order of magnitude faster than the iteration-based algorithm. For $64^3$

voxels of a 3D tomogram as a reference data point, it took only 0.159 sec for DeepRegularizer to improve the resolution of the input tomogram while the iterative TV took 9.13 sec. In fact, the inference time difference between the two methods increased for a larger tomogram, e.g. DeepReguarizer and the iterative TV took 2.27 sec and 24.497, respectively for a $320^2 \times 64$ tomogram. The speed enhancement factor between the two methods decreases because the inference time for our framework is constrained by the number of tomogram patches. Further improvement in speed is expected by optimizing the stitching algorithm.

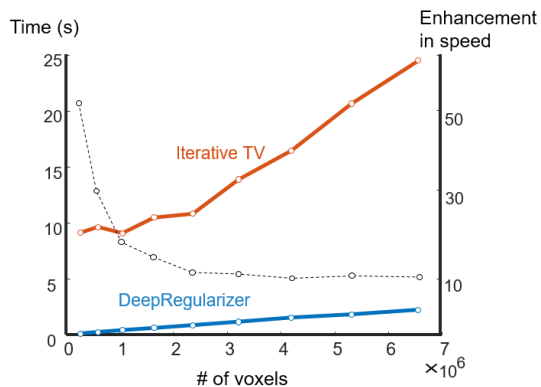

Fig. 6. Comparing the inference time between deep learning and iterative TV. Our deep learning approach can be more than an order of magnitude faster than the iterative TV regularization. The black dotted line indicates the enhancement factor in speed between the two approaches, which decreases as the number of input voxels increases.

## IV. DISCUSSION AND CONCLUSION

In summary, we have proposed and experimentally demonstrated a data-driven approach that rapidly enhances the resolution of 3D RI maps in ODT. The 3D deep neural network, optimized by the neural architecture search, learns a statistical transformation between image pairs (raw 3D tomogram and parameter-tuned TV regularized tomogram) and fills in the missing cone, treating severe $k_z$ artefacts within a few seconds. We believe that the inference time could decrease further with the additional optimization of the network architecture. The experimental performance and its generalizability were verified by reconstructing *S. pyogenes* and OCI-AML3 cells. Additionally, we tested our method on various suspension (MV4-11 and monocyte) and adherent cell types (H292 and NIH3T3). We envision that this framework can be widely employed to approximate computationally intensive optimization problems, including the missing cone problem in other tomographic imaging modalities [53-57].

With minimal efforts in optimizing the network, we trained a robust 3D computational regularizer with excellent experimental capability, generalizability, and to be highly flexible and applicable to other regularization problems. Unlike the conventional regularization algorithm, the proposed network exploits automated machine learning and the parameters in the training process are easily tunable. Most neural parameters in our network can be automatically optimized and we attempted to change a few hyperparameters, such as batch size, for further tuning. In addition, once the ground truth dataset is generated based on human perception, it is rather straightforward to design a loss function that targets the ground truth in the process of training. An immediate application of this approach would be a deep network that replaces other computationally heavy regularizers and exploits sparsity or edge-preserving smoothness.

One of the main quests for the existing 'data-hungry' approaches is to enhance generalizability. Although DeepRegularizer has proved its excellent generalization across different samples, we could further expand the limit of our model by leveraging the extensive range of a large dataset; however, this is not always feasible in practice. Alternatively, various techniques in deep learning, such as transfer learning [58], which uses an already trained model to effectively create and optimize a model for other purposes, and deep image prior, learning from the structure of a neural network rather than a training dataset [59], could be employed to mitigate the need for a large dataset.

Several future works can be considered to extend the approach presented herein. First, it would be worthwhile to implement a TV regularizing network that maps between Fourier domains, i.e. instead of considering real-valued RIs, the complex-valued network [60] approximating a function that translates between two different scattering potentials in Fourier domain can be considered. Second, an end-to-end network that transforms measured interferograms to a regularized tomogram can be implemented [61]. Though design and optimization of a many-to-one neural network may require more sophisticated architecture than a one-to-one network, it would enable faster 3D reconstruction. Finally, an interesting extension of this work would be a deep neural network that finds optimal parameters for TV regularization. Instead of manually preparing a good paired dataset for deep learning, we may include the algorithmic parameters as meta-parameters in deep network training [62].

## APPENDIX

### A. Parameter Dependency of the Total Variation Regularization with Non-negativity Constraint

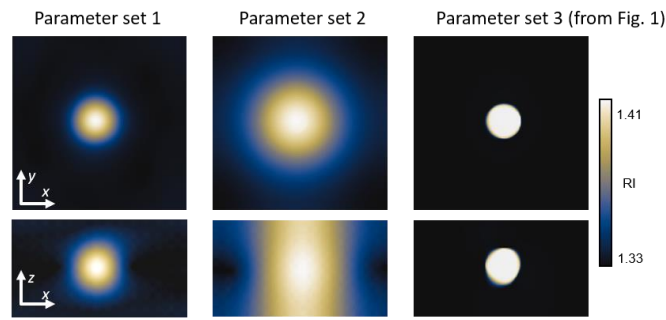

Fig. S1. Regularization with different parameters. For the parameter set (N, M, μ, τ, γ), three sets are used to regularize the tomograms of SiO2 beads, and the lateral and axial cross-sections of the bead tomogram are visualized. First, as for inappropriate regularization examples, (2, 400, 2, 2, 1) and (2, 400, 10, 2, 1) are used for the set 1 and set 2. These two regularized tomograms significantly blur out. On the other hand, proper regularization using the parameter set (2, 400, 10, 10, 1) leads to the accurate reconstruction of the bead tomogram with a sharp boundary in both lateral and axial view.

## B. Derivation of the Iterative Regularization Algorithm using Variable Splitting

The equation (3) from the main text can be solved via the variable splitting. With the replacement $w \leftarrow f, d \leftarrow \nabla f$ the split formulation of (3) then becomes

$$\min_{f,d_x,d_y,d_z,w} \|(d_x, d_y, d_z)\|_2 + N(w \geq 0)$$
$$+ |\frac{\mu}{2}| Af - g \|_2^2 + \frac{\lambda}{2}(\| d_x - \nabla_x f - b_x \|_2^2$$
$$+ \| d_y - \nabla_y f - b_y \|_2^2 + \| d_z - \nabla_z f - b_z \|_2^2)$$
$$+ \frac{\gamma}{2}(w - f - b_w)^2 \quad (1.1)$$

To find optimal value of f, we solve the following subproblem.

$$f^{k+1} = \min_f \frac{\mu}{2} | Af - g \|_2^2$$
$$+ \frac{\lambda}{2}(\| d_x - \nabla_x f - b_x \|_2^2$$
$$+ \| d_y - \nabla_y f - b_y \|_2^2 + \| d_z - \nabla_z f - b_z \|_2^2)$$
$$+ \frac{\gamma}{2}(w - f - b_w)^2 \quad (1.2)$$

Since this sub-problem is differentiable, we can obtain optimality condition for $f^{k+1}$ by simply taking a derivative with respect to $f$.

$$[\mu A^T A - \lambda(\nabla_x^T \nabla_x + \nabla_y^T \nabla_y + \nabla_z^T \nabla_z) + \gamma I] f^{k+1}$$
$$= [\mu A^T g + \lambda(\nabla_x^T (d_x - b_x) + \nabla_y^T (d_y - b_y)$$
$$+ \nabla_z^T (d_z - b_z)) + \gamma(w - b_w)] = rhs^k \quad (1.3)$$

The optimal $f^{k+1}$ can be iteratively solved as follows:

$$f^{k+1} = [\mu A^T A + \lambda \Delta + \gamma I]^{-1} rhs^k \quad (1.4)$$

where $\Delta = -(\nabla_x^T \nabla_x + \nabla_y^T \nabla_y + \nabla_z^T \nabla_z)$.

The remaining elements in the optimization problem (1.1), split-variables $d_x, d_y, d_z, w$, can be solved via shrinkage operators because the elements are not coupled.

$$d_x^{k+1} = \max(s^k - \frac{1}{\lambda}, 0) \frac{\nabla_x f^{k+1} + b_x^k}{s^k} \quad (1.5)$$

$$d_y^{k+1} = \max(s^k - \frac{1}{\lambda}, 0) \frac{\nabla_y f^{k+1} + b_y^k}{s^k} \quad (1.6)$$

$$d_z^{k+1} = \max(s^k - \frac{1}{\lambda}, 0) \frac{\nabla_z f^{k+1} + b_z^k}{s^k} \quad (1.7)$$

$$w^{k+1} = \max(| f^{k+1} + b_w^k | - \frac{1}{\gamma}, 0) \frac{f^{k+1} + b_w^k}{| f^{k+1} + b_w^k |} \quad (1.8)$$

$$b_x^{k+1} = b_x^k + (\nabla_x^T f^{k+1} - d_x^{k+1}) \quad (1.9)$$
$$b_y^{k+1} = b_y^k + (\nabla_y^T f^{k+1} - d_y^{k+1}) \quad (1.10)$$
$$b_z^{k+1} = b_z^k + (\nabla_z^T f^{k+1} - d_z^{k+1}) \quad (1.11)$$
$$b_w^{k+1} = b_x^k + (\nabla_x^T f^{k+1} - w^{k+1}) \quad (1.12)$$

Finally, we implement the iterative regularization algorithm by repetitively seeking each solution of (1.4), (1.5), (1.6), (1.7), (1.8), (1.9), (1.10), (1.11), and (1.12) for the inner iteration, and (4) from the main text for the outer iteration.

## C. Generalization test of DeepRegularizer

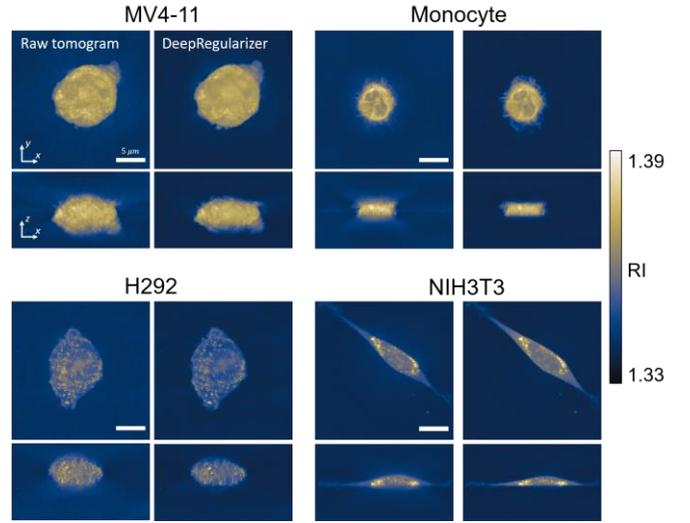

Fig. S2. Generalization test of DeepRegularizer. Our deep learning model trained with the OCI-AML3 dataset is assessed on MV4-11, Monocyte, H292, and NIH3T3.

For additional generalization tests, we regularized unseen cell tomograms, including MV4-11, peripheral blood monocyte, H292, and NIH3T3, using our original model trained with OCI-AML3 dataset, as depicted in Fig. S2. The unwanted noises and elongated tail artifacts in each transverse and axial view was successfully removed by the DeepRegularizer. It is noteworthy that some artifacts of horizontal stripes remained in the background of the H292 cell and NIH3T3. We postulate that the model trained with the suspension cells (OCI-AML3) could have limited generalizability on the relatively flat adherent cells, such as H292 and NIH3T3, which might be improved by adding various cell types in the training set.

### D. Smaller 3D U-net inference

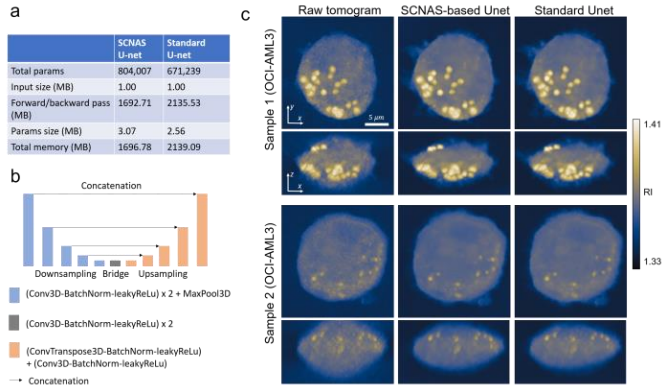

Fig. S3. Comparison of our architecture designed with SCNAS and a standard U-net. (a) Memory used by SCNAS U-net and standard U-net. (b) Architecture of the standard U-net. (c) Regularizing performance of two different U-net on unseen OCI-AML3 data.

We also conducted a training with a standard U-net that does not include stochastic upsampling/downsampling cells optimized by the scalable neural architecture search (SCNAS). This smaller U-net consists of downsampling and upsampling path where 3D convolution, batch normalization and leaky ReLu alternately repeated along with concatenating path (Fig. S3a). For reasonable comparison, we trained the new network with the same hyper-parameters as used in the original model for the same period of training time.

In Fig. S3b, the regularization of our original model and the smaller U-net is assessed on the unseen OCI-AML3 dataset, which shows a comparable performance. While our network optimized by SCNAS can be easily implemented to perform optimal performance on image-to-image translation tasks, the standard-type 3D U-net with smaller number of learnable parameters could still be used per the purpose.

### ACKNOWLEDGMENT

We thank Joowon Lim (EPFL) for his advice on regularization algorithms.

### CONFLICT OF INTEREST

D.R., H.C., H.M., and Y.P have financial interests in Tomocube Inc., a company that commercializes optical diffraction tomography and quantitative phase imaging instruments and is one of the sponsors of the work.

### REFERENCES

[1] Y. Park, C. Depeursinge, and G. Popescu, "Quantitative phase imaging in biomedicine," *Nat. Photonics,* vol. 12, no. 10, p. 578, 2018.
[2] V. Lauer, "New approach to optical diffraction tomography yielding a vector equation of diffraction tomography and a novel tomographic microscope," *J. Microsc.,* vol. 205, no. 2, pp. 165-176, 2002.
[3] M. Born and E. Wolf, *Principles of optics: electromagnetic theory of propagation, interference and diffraction of light*. Elsevier, 2013.
[4] K. Lee *et al.*, "Quantitative phase imaging techniques for the study of cell pathophysiology: from principles to applications," *Sensors,* vol. 13, no. 4, pp. 4170-4191, 2013.
[5] F. Charrière *et al.*, "Cell refractive index tomography by digital holographic microscopy," *Opt. Lett.,* vol. 31, no. 2, pp. 178-180, 2006.
[6] W. Choi *et al.*, "Tomographic phase microscopy," *Nat. Methods,* vol. 4, no. 9, pp. 717-719, 2007.
[7] S. Park *et al.*, "Label-Free Tomographic Imaging of Lipid Droplets in Foam Cells for Machine-Learning-Assisted Therapeutic Evaluation of Targeted Nanodrugs," *ACS Nano,* 2020.
[8] J. Jung *et al.*, "Label-free non-invasive quantitative measurement of lipid contents in individual microalgal cells using refractive index tomography," *Sci. Rep.,* vol. 8, no. 1, p. 6524, 2018.
[9] S. Lee, H. Park, K. Kim, Y. Sohn, S. Jang, and Y. Park, "Refractive index tomograms and dynamic membrane fluctuations of red blood cells from patients with diabetes mellitus," *Sci. Rep.,* vol. 7, no. 1, pp. 1-11, 2017.
[10] D. Kim *et al.*, "Label-free high-resolution 3-D imaging of gold nanoparticles inside live cells using optical diffraction tomography," *Methods,* vol. 136, pp. 160-167, 2018.
[11] K. Tam and V. Perez-Mendez, "Tomographical imaging with limited-angle input," *J. Opt. Soc. Am.,* vol. 71, no. 5, pp. 582-592, 1981.
[12] C. Park, S. Shin, and Y. Park, "Generalized quantification of three-dimensional resolution in optical diffraction tomography using the projection of maximal spatial bandwidths," *J. opt. Soc. Am. A,* vol. 35, no. 11, pp. 1891-1898, 2018.
[13] P. Müller, M. Schürmann, C. J. Chan, and J. Guck, "Single-cell diffraction tomography with optofluidic rotation about a tilted axis," in *Optical Trapping and Optical Micromanipulation XII*, 2015, vol. 9548, p. 95480U: International Society for Optics and Photonics.
[14] F. Merola *et al.*, "Tomographic flow cytometry by digital holography," *Light Sci. Appl.,* vol. 6, no. 4, p. e16241, 2017.
[15] B. Simon *et al.*, "Tomographic diffractive microscopy with isotropic resolution," *Optica,* vol. 4, no. 4, pp. 460-463, 2017.
[16] S. Shin *et al.*, "Enhancement of optical resolution in three-dimensional refractive-index tomograms of biological samples by employing micromirror-embedded coverslips," *Lab Chip,* vol. 18, no. 22, pp. 3484-3491, 2018.
[17] J. Lim *et al.*, "Comparative study of iterative reconstruction algorithms for missing cone problems in optical diffraction tomography," *Opt. Express,* vol. 23, no. 13, pp. 16933-16948, 2015.
[18] Y. Sung, W. Choi, N. Lue, R. R. Dasari, and Z. Yaqoob, "Stain-free quantification of chromosomes in live cells using regularized tomographic phase microscopy," *PLoS One,* vol. 7, no. 11, p. e49502, 2012.
[19] A. H. Delaney and Y. Bresler, "Globally convergent edge-preserving regularized reconstruction: an application to limited-angle tomography," *IEEE Trans. Image Process.,* vol. 7, no. 2, pp. 204-221, 1998.
[20] Y. Sung and R. R. Dasari, "Deterministic regularization of three-dimensional optical diffraction tomography," *J. Opt. Soc. Am. A,* vol. 28, no. 8, pp. 1554-1561, 2011.
[21] B. Goris, W. Van den Broek, K. J. Batenburg, H. H. Mezerji, and S. Bals, "Electron tomography based on a total variation minimization reconstruction technique," *Ultramicroscopy,* vol. 113, pp. 120-130, 2012.
[22] W. Krauze, P. Makowski, M. Kujawińska, and A. Kuś, "Generalized total variation iterative constraint strategy in limited angle optical diffraction tomography," *Opt. Express,* vol. 24, no. 5, pp. 4924-4936, 2016.
[23] T. Goldstein and S. Osher, "The split Bregman method for L1-regularized problems," *SIAM J. IMAGING SCI.,* vol. 2, no. 2, pp. 323-343, 2009.
[24] U. S. Kamilov *et al.*, "Optical tomographic image reconstruction based on beam propagation and sparse regularization," *IEEE Trans. Comput. Imaging,* vol. 2, no. 1, pp. 59-70, 2016.
[25] S. Chowdhury *et al.*, "High-resolution 3D refractive index microscopy of multiple-scattering samples from intensity images," *Optica,* vol. 6, no. 9, pp. 1211-1219, 2019.
[26] J. Lim, A. B. Ayoub, E. E. Antoine, and D. Psaltis, "High-fidelity optical diffraction tomography of multiple scattering samples," *Light Sci. Appl.,* vol. 8, no. 1, pp. 1-12, 2019.
[27] T.-a. Pham, E. Soubies, A. Ayoub, J. Lim, D. Psaltis, and M. Unser, "Three-Dimensional Optical Diffraction Tomography with Lippmann-Schwinger Model," *IEEE Trans. Comput. Imaging,* 2020.

[28] P. Memmolo et al., "How holographic imaging can improve machine learning," in *Multimodal Sensing: Technologies and Applications*, 2019, vol. 11059, p. 1105908: International Society for Optics and Photonics.

[29] M. T. McCann, K. H. Jin, and M. Unser, "Convolutional neural networks for inverse problems in imaging: A review," *IEEE Signal Process. Mag.,* vol. 34, no. 6, pp. 85-95, 2017.

[30] Y. Jo et al., "Quantitative phase imaging and artificial intelligence: a review," *IEEE J. Sel. Top. Quantum Electron.,* vol. 25, no. 1, pp. 1-14, 2018.

[31] G. Barbastathis, A. Ozcan, and G. Situ, "On the use of deep learning for computational imaging," *Optica,* vol. 6, no. 8, pp. 921-943, 2019.

[32] Y. Rivenson, Y. Wu, and A. Ozcan, "Deep learning in holography and coherent imaging," *Light Sci. Appl.,* vol. 8, no. 1, pp. 1-8, 2019.

[33] E. Nehme, L. E. Weiss, T. Michaeli, and Y. Shechtman, "Deep-STORM: super-resolution single-molecule microscopy by deep learning," *Optica,* vol. 5, no. 4, pp. 458-464, 2018.

[34] W. Ouyang, A. Aristov, M. Lelek, X. Hao, and C. Zimmer, "Deep learning massively accelerates super-resolution localization microscopy," *Nat. Biotechnol.,* vol. 36, no. 5, p. 460, 2018.

[35] B. Rahmani, D. Loterie, G. Konstantinou, D. Psaltis, and C. Moser, "Multimode optical fiber transmission with a deep learning network," *Light Sci. Appl.,* vol. 7, no. 1, pp. 1-11, 2018.

[36] M. Lyu, H. Wang, G. Li, and G. Situ, "Exploit imaging through opaque wall via deep learning," *Preprint at https://arxiv.org/abs/1708.07881s,* 2017.

[37] S. Li, M. Deng, J. Lee, A. Sinha, and G. Barbastathis, "Imaging through glass diffusers using densely connected convolutional networks," *Optica,* vol. 5, no. 7, pp. 803-813, 2018.

[38] J. Shao, J. Zhang, X. Huang, R. Liang, and K. Barnard, "Fiber bundle image restoration using deep learning," *Opt. Lett.,* vol. 44, no. 5, pp. 1080-1083, 2019.

[39] F. Pan, B. Dong, W. Xiao, and P. Ferraro, "Stitching sub-aperture in digital holography based on machine learning," *Opt. Express,* vol. 28, no. 5, pp. 6537-6551, 2020.

[40] Y. Rivenson, Y. Zhang, H. Günaydın, D. Teng, and A. Ozcan, "Phase recovery and holographic image reconstruction using deep learning in neural networks," *Light Sci. Appl.,* vol. 7, no. 2, p. 17141, 2018.

[41] T. Nguyen, Y. Xue, Y. Li, L. Tian, and G. Nehmetallah, "Deep learning approach for Fourier ptychography microscopy," *Opt. Express,* vol. 26, no. 20, pp. 26470-26484, 2018.

[42] W. Jeon, W. Jeong, K. Son, and H. Yang, "Speckle noise reduction for digital holographic images using multi-scale convolutional neural networks," *Opt. Lett.,* vol. 43, no. 17, pp. 4240-4243, 2018.

[43] G. Choi et al., "Cycle-consistent deep learning approach to coherent noise reduction in optical diffraction tomography," *Opt. Express,* vol. 27, no. 4, pp. 4927-4943, 2019.

[44] M. E. Kandel et al., "PICS: Phase Imaging with Computational Specificity," *Preprint at https://arxiv.org/abs/2002.08361,* 2020.

[45] E. M. Christiansen et al., "In silico labeling: predicting fluorescent labels in unlabeled images," *Cell,* vol. 173, no. 3, pp. 792-803. e19, 2018.

[46] Y. Rivenson, T. Liu, Z. Wei, Y. Zhang, K. de Haan, and A. Ozcan, "PhaseStain: the digital staining of label-free quantitative phase microscopy images using deep learning," *Light Sci. Appl.,* vol. 8, no. 1, p. 23, 2019.

[47] F. Wang, H. Wang, H. Wang, G. Li, and G. Situ, "Learning from simulation: An end-to-end deep-learning approach for computational ghost imaging," *Opt. Express,* vol. 27, no. 18, pp. 25560-25572, 2019.

[48] S. Kim et al., "Scalable Neural Architecture Search for 3D Medical Image Segmentation," *Preprint at https://arxiv.org/abs/1906.05956,* 2019.

[49] S. Kim et al., "Scalable Neural Architecture Search for 3D Medical Image Segmentation," *Preprint at https://arxiv.org/abs/1906.05956,* 2019.

[50] E. Jang, S. Gu, and B. Poole, "Categorical reparameterization with gumbel-softmax," *Preprint at https://arxiv.org/abs/1611.01144,* 2016.

[51] E. Cuche, P. Marquet, and C. Depeursinge, "Spatial filtering for zero-order and twin-image elimination in digital off-axis holography," *Appl. Opt.,* vol. 39, no. 23, pp. 4070-4075, 2000.

[52] E. Wolf, "Three-dimensional structure determination of semi-transparent objects from holographic data," *Opt. Commun.,* vol. 1, no. 4, pp. 153-156, 1969.

[53] M. Lustig, D. Donoho, and J. M. J. M. R. i. M. A. O. J. o. t. I. S. f. M. R. i. M. Pauly, "Sparse MRI: The application of compressed sensing for rapid MR imaging," vol. 58, no. 6, pp. 1182-1195, 2007.

[54] S. Bartolac, R. Clackdoyle, F. Noo, J. Siewerdsen, D. Moseley, and D. J. M. p. Jaffray, "A local shift-variant Fourier model and experimental validation of circular cone-beam computed tomography artifacts," vol. 36, no. 2, pp. 500-512, 2009.

[55] Z. Wang et al., "Spatial light interference tomography (SLIT)," vol. 19, no. 21, pp. 19907-19918, 2011.

[56] M. Fauver et al., "Three-dimensional imaging of single isolated cell nuclei using optical projection tomography," vol. 13, no. 11, pp. 4210-4223, 2005.

[57] Z. Turani et al., "Optical radiomic signatures derived from optical coherence tomography images improve identification of melanoma," vol. 79, no. 8, pp. 2021-2030, 2019.

[58] S. J. Pan and Q. Yang, "A survey on transfer learning," *IEEE Trans. Knowl. Data Eng.,* vol. 22, no. 10, pp. 1345-1359, 2009.

[59] D. Ulyanov, A. Vedaldi, and V. Lempitsky, "Deep image prior," in *Proc. IEEE Comput. Soc. Conf. Comput. Vis. Pattern Recognit.*, 2018, pp. 9446-9454.

[60] C. Trabelsi et al., "Deep complex networks," *Preprint at https://arxiv.org/abs/1705.09792,* 2017.

[61] A. Goy, G. Rughoobur, S. Li, K. Arthur, A. I. Akinwande, and G. Barbastathis, "High-resolution limited-angle phase tomography of dense layered objects using deep neural networks," *Proc. Natl. Acad. Sci. U.S.A,* vol. 116, no. 40, pp. 19848-19856, 2019.

[62] Y. Wu et al., "Three-dimensional virtual refocusing of fluorescence microscopy images using deep learning," *Nat. Methods,* vol. 16, no. 12, pp. 1323-1331, 2019.